\documentclass[aps,twocolumn,prl]{revtex4}

\usepackage{graphicx}
\usepackage{amsmath}
\usepackage{amssymb}
\usepackage{units}
\usepackage{times}

\begin{document}

%
%

\title{Electronic decoupling of an epitaxial graphene monolayer by gold intercalation}
\author{Isabella\ Gierz$^1$}
\email[Corresponding author; electronic address:\
]{i.gierz@fkf.mpg.de}
\author{Takayuki Suzuki$^1$}
\author{Dong Su Lee$^1$}
\author{Benjamin Krauss$^1$}
\author{Christian Riedl$^1$}
\author{Ulrich Starke$^1$}
\author{Hartmut H\"ochst$^2$}
\author{Jurgen H. Smet$^1$}
\author{Christian R. Ast$^1$}
\author{Klaus Kern$^{1,3}$}
\affiliation{$^1$ Max-Planck-Institut f\"ur Festk\"orperforschung, D-70569 Stuttgart, Germany\\
$^2$ Synchrotron Radiation Center, University of Wisconsin-Madison, Stoughton, WI, 53589, USA\\
$^3$ Institut de Physique de la Mati{\`e}re Condens{\'e}e, Ecole Polytechnique F{\'e}d{\'e}rale de Lausanne, CH-1015 Lausanne, Switzerland}

\date{\today}


\begin{abstract}
The application of graphene in electronic devices
requires large scale epitaxial growth. The presence of the substrate, however, usually reduces the
charge carrier mobility considerably. We show that it is possible
to decouple the partially sp$^3$-hybridized first graphitic layer
formed on the Si-terminated face of silicon carbide from the substrate by
gold intercalation, leading to a completely sp$^2$-hybridized
graphene layer with improved electronic properties.
\end{abstract}

\maketitle


Electrons in graphene --- sp$^2$-bonded carbon atoms arranged
in a honeycomb lattice --- behave like massless Dirac particles
and exhibit an extremely high carrier mobility \cite{Geim1}. So
far, the only feasible route towards large scale production of
graphene is epitaxial growth on a substrate. The presence of the
substrate will, however, influence the electronic
properties of the graphene layer. To preserve its unique
properties it is desirable to decouple the graphene layer from the
substrate. Here we present a new approach for the growth
of highly decoupled epitaxial graphene on a silicon carbide
substrate. By decoupling the strongly interacting, partially
sp$^3$-hybridized first graphitic layer (commonly referred to as
zero layer (ZL) \cite{Bostwick0}) from the SiC(0001) substrate by gold
intercalation, we obtain a completely sp$^2$-hybridized graphene
layer with improved electronic properties as confirmed by
angle-resolved photoemission spectroscopy (ARPES), scanning
tunneling microscopy (STM) and Raman spectroscopy.

There are essentially two ways for large scale epitaxial growth of
graphene on a substrate: by cracking organic molecules on
catalytic metal surfaces
\cite{Preobrajenski1,Sasaki1,Pletikosic1,Varykhalov1,Sutter1} or
by thermal graphitization of SiC
\cite{Bostwick0,Bostwick,Ohta,Zhou,Berger1}. Unfortunately, the
presence of the substrate alters the electronic properties of the
graphene layer on the surface and reduces the carrier mobility.
Even though it has been shown that the
graphene layer can be decoupled from a metallic substrate
\cite{Shikin1,Varykhalov1,Farias1,Dedkov1} the system remains unsuitable for
device applications. This problem can be
solved by decoupling the graphene layer from a semiconducting SiC
substrate \cite{Riedl1}.


\begin{figure*}
  \includegraphics[width = 2\columnwidth]{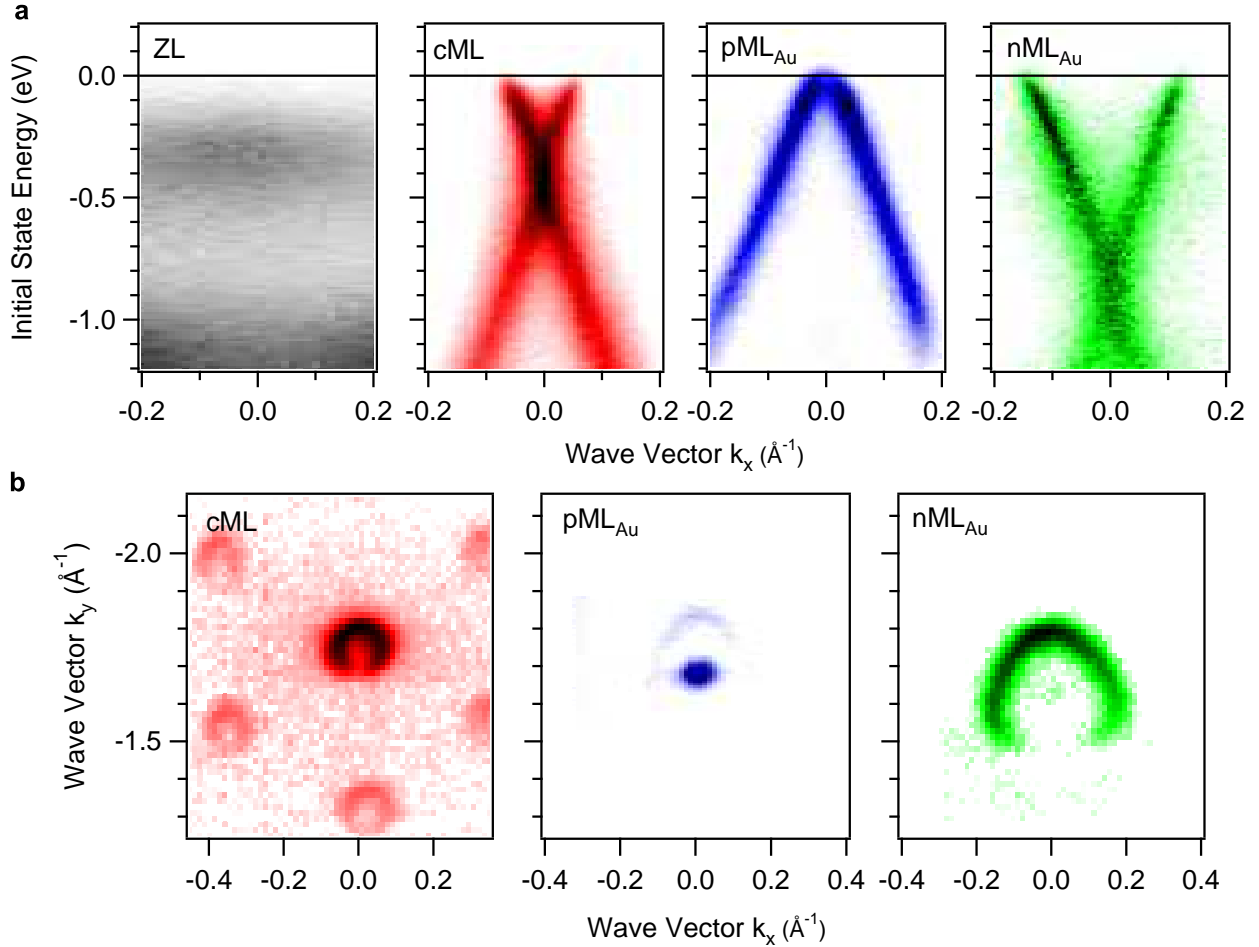}
  \caption{(color online) Comparison of ARPES data for conventional graphene
  on SiC and graphene intercalated with Au: panel a) shows the
  band structure measured in the direction perpendicular to the $\overline{\Gamma \mbox{K}}$ direction near the $\overline{\mbox{K}}$-point of the surface Brillouin zone of the zero layer (black), the conventional graphene monolayer (red), the p-doped graphene monolayer intercalated with gold (blue) and the n-doped graphene monolayer intercalated with gold (green) together with the corresponding Fermi surfaces in panel b). The Fermi surfaces are plotted on a logarithmic color scale to enhance weak features. $k_x$ is perpendicular to the $\overline{\Gamma \mbox{K}}$ direction, $k_y$ is along the $\overline{\Gamma \mbox{K}}$ direction. The Fermi surface for the p-doped graphene monolayer shows a weak contribution of the n-doped phase due to an inhomogeneous Au coverage on the sample.}
  \label{figure1}
\end{figure*}

On both the silicon and the carbon terminated face of a SiC substrate, graphene is commonly grown by thermal
graphitization in ultra high vacuum (UHV). When annealing the
substrate at elevated temperatures Si atoms leave the surface
whereas the C atoms remain and form carbon layers. On SiC(000$\overline{1}$), the so-called C-face, the weak graphene-to-substrate interaction results in the growth of rotationally disordered multilayer graphene and a precise thickness control becomes difficult \cite{Emtsev}. On the other hand, the rotational disorder decouples the graphene layers so that the transport properties resemble those of isolated graphene sheets with room temperature mobilities in excess of 200,000\,cm$^2$/Vs \cite{Sprinkle}.

On SiC(0001), i. e. the Si-face, the comparatively strong
graphene-to-substrate interaction results in uniform, long-range
ordered layer-by-layer growth. The first carbon
layer (=ZL) grown on the Si-face is partially sp$^3$-hybridized to
the substrate, which means that on a ZL no $\pi$-bands can develop and it has no graphene properties.
This can be seen in the first panel of Fig.\ \ref{figure1} a),
where the experimental band structure of the ZL (black) measured by ARPES near the
$\overline{\mbox{K}}$-point of the surface Brillouin zone is shown. The ZL lacks the linear
dispersion typical for graphene $\pi$-bands. Its band structure consists of
two non-dispersing bands at about $-0.3$\,eV and $-1.2$\,eV
initial state energy. In addition, the ZL forms a
($6\sqrt{3}\times6\sqrt{3}$)R30$^{\circ}$ reconstruction with
respect to the SiC substrate \cite{Bostwick0,Emtsev,Varchon}.

Further graphitization leads to the growth of a completely
sp$^2$-hybridized graphene layer, for which the ZL acts as a
buffer layer. The band structure of this
``conventionally'' grown graphene monolayer (cML) near the $\overline{\mbox{K}}$-point 
is shown in the second panel of Fig.\ \ref{figure1}
a). The cML is influenced considerably by the underlying SiC substrate.
It is n-doped with the crossing point of the
two linear bands (Dirac point) at $E_D=-420$\,meV due to charge
transfer from the substrate \cite{Bostwick,Ohta,Zhou,Riedl1}. Furthermore, the possibility of a band gap opening has been suggested \cite{Zhou} and explained theoretically in connection with the formation of midgap states \cite{Kim}. In addition to that, the strong substrate influence reduces
the carrier mobility considerably \cite{Robinson}.
The ($6\sqrt{3}\times6\sqrt{3}$)R30$^{\circ}$ reconstruction of the ZL
diffracts the outgoing photoelectrons giving rise to the formation of replica bands \cite{Bostwick0}. This is nicely seen in the measured Fermi surface of the cML around $\overline{\mbox{K}}$
in the left panel of Fig.\ \ref{figure1} b). The size of the Fermi
surface is determined by the charge carrier density $n=k_F^2/\pi$,
where $k_F$ is the Fermi wave vector with respect to the $\overline{\mbox{K}}$-point. The values are summarized in
Table \ref{table}.

\begin{table}
\caption{{\bf Characteristic parameters for cML, pML$_{\text{Au}}$ and nML$_{\text{Au}}$}
determined from the photoemission experiments.}
\begin{center}
\begin{tabular}{|c||c|c|c|c|}
\hline
 & cML & pML$_{\text{Au}}$ & nML$_{\text{Au}}$ \\
\hline
\hline
Au coverage (ML) & 0 & 1 & 1/3 \\
\hline
Au-Si 4f$_{5/2}$ (eV)& & 88.20 & 89.05\\
Au-Si 4f$_{7/2}$ (eV)& & 84.54 & 85.41\\
\hline
Au-Au 4f$_{5/2}$ (eV)&  & 87.82 & 88.32  \\
Au-Au 4f$_{7/2}$ (eV)&  & 84.15 & 84.68 \\
\hline
charge carrier & 1$\times$10$^{13}$ & 7$\times$10$^{11}$ & 5$\times$10$^{13}$ \\
density (cm$^{-2}$)& electrons & holes & electrons \\
\hline
Dirac point (meV) & $-420$ & $+100$ & $-850$ \\
\hline
\end{tabular}
\end{center}
\label{table}
\end{table}

To reduce the influence of the substrate we developed a new method
for the epitaxial growth of graphene on the Si-face of SiC. We
start with the preparation of the ZL exploiting the strong
substrate influence for uniform growth. On top of the ZL, we
deposit Au atoms at room temperature. After subsequent annealing
of the sample at 800$^{\circ}$C the linear dispersion typical for graphene appears.
Depending on the gold coverage (about one third or one
monolayer, respectively), either a strongly n-doped (nML$_{\text{Au}}$) or a p-doped
(pML$_{\text{Au}}$) graphene layer is formed. The band
structures for the pML$_{\text{Au}}$ and the nML$_{\text{Au}}$ are
compared in Fig.\ \ref{figure1} a). In contrast to the ZL, both the
pML$_{\text{Au}}$ (blue) and the nML$_{\text{Au}}$ (green) clearly
show two linearly dispersing $\pi$-bands. The Dirac point for the
pML$_{\text{Au}}$ is about 100\,meV above the Fermi level. This
band structure looks similar to the one reported in Ref.\
\cite{Gierz}. However, there the graphene monolayer was prepared by
depositing Au directly on a cML and not on a ZL as in this work. For the
nML$_{\text{Au}}$ the bands cross at about $-850$\,meV. The band structure of the cML is a superposition of the band structure of the underlying ZL and the graphene monolayer. Both
pML$_{\text{Au}}$ and nML$_{\text{Au}}$, however, are formed directly from the ZL. There is no additional carbon layer between the graphene layer and the substrate. Therefore, the band structure around the $\overline{\mbox{K}}$-point is given by the pML$_{\text{Au}}$ and nML$_{\text{Au}}$ alone. The
charge carrier densities deduced from the size of the Fermi
surface (see middle and right panel of Fig.\ \ref{figure1} b) are
listed in Table \ref{table}.

\begin{figure*}
  \includegraphics[width = 2\columnwidth]{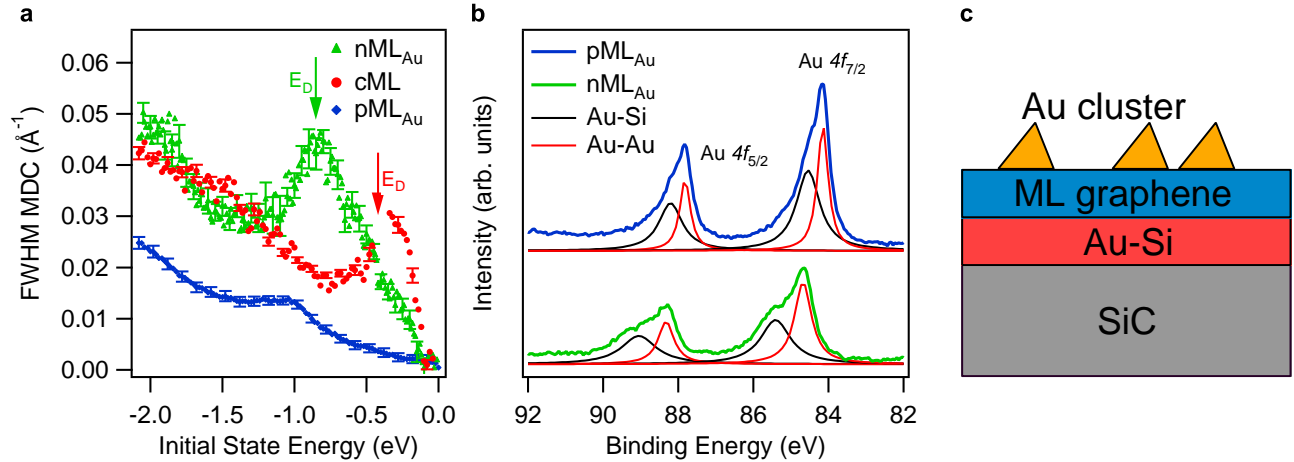}
  \caption{(color online) Linewidth analysis, Au {\it 4f} core level spectra and
  schematic: Panel a) shows the full width at half maximum (FWHM)
  of momentum distribution curves obtained from Fig.\ \ref{figure1} a)
  for the conventional graphene monolayer (red), the p-doped graphene monolayer intercalated with Au (blue) and the n-doped graphene monolayer intercalated with Au (green).
  A constant background was
  subtracted from the data so that the plotted linewidth is determined
  by electron-phonon, electron-plasmon and electron-electron scattering
  alone. Panel b) shows the Au $4f$ core level spectra recorded with an
  incident photon energy of 150\,eV for the p-doped monolayer (blue) and the
  n-doped monolayer (green). The core level spectra indicate the presence of
  Au-Si bonds (black lines) for both the p- and the n-doped monolayer
  which is consistent with the structural model shown in panel c).}
  \label{figure2}
\end{figure*}

Comparing the Fermi surfaces for the cML (red), the
pML$_{\text{Au}}$ (blue), and the nML$_{\text{Au}}$ (green) in
Fig.\ \ref{figure1} b), the most striking difference is the absence
of replica bands for the pML$_{\text{Au}}$ and nML$_{\text{Au}}$. Even on the logarithmic color scale of Fig.\ \ref{figure1} b) the replica bands are invisible, indicating a reduced influence of the
($6\sqrt{3}\times6\sqrt{3}$)R30$^{\circ}$ reconstruction. Low
energy electron diffraction (LEED) images (shown in the
EPAPS) reveal a strong decrease of the
intensity for spots related to the
($6\sqrt{3}\times6\sqrt{3}$)R30$^{\circ}$ reconstruction for the
pML$_{\text{Au}}$ as compared to the graphene-related spots. For
the nML$_{\text{Au}}$, however, the ($6\sqrt{3}\times6\sqrt{3}$)R30$^{\circ}$ spots have a similar
intensity as for the cML. We conclude that only the pML$_{\text{Au}}$
is less influenced by the underlying substrate. We attribute this to an increased graphene-to-substrate distance as will be discussed later in this paper. 

To analyze the band structure in more detail and gain access to
the relevant scattering mechanisms we determined the full width at
half maximum (FWHM) of the bands by fitting momentum distribution
curves (MDCs) along the $\overline{\Gamma{\text{K}}}$ direction with Lorentzian lineshapes and a constant
background. The FWHM as a function of the initial state energy for
the cML (red), the pML$_{\text{Au}}$ (blue) and the
nML$_{\text{Au}}$ (green) are shown in Fig.\ \ref{figure2} a). From the data in Fig.\ \ref{figure2} a) a constant offset of 0.023\,\AA$^{-1}$ (cML), 0.027\,\AA$^{-1}$ (pML$_{\text{Au}}$), and
0.041\,\AA$^{-1}$ (nML$_{\text{Au}}$) has been subtracted. For both cML and pML$_{\text{Au}}$ this offset is mainly determined by the experimental resolution (see EPAPS). For the nML$_{\text{Au}}$, however, the linewidth offset is significantly larger than the limit set by the experimental resolution. In this case the offset is determined by impurity scattering which gives a constant contribution to the linewidth at all energies. 

There are three main contributions to
the quasiparticle lifetime in graphene \cite{Bostwick0,Bostwick}.
The increase in linewidth around 200\,meV is caused by electron-phonon coupling which depends on the size of the Fermi surface. Therefore, its influence is largest for strongly n-doped graphene (nML$_{\text{Au}}$). The pronounced maximum near the Dirac point is attributed to
electron-plasmon scattering. The third contribution to the line
width is electron-electron scattering, which has been found to be
proportional to $|E-E_F|^{\alpha}$, where $1<\alpha<2$ \cite{Bostwick}. The FWHM
for our cML is in good agreement with the data reported in Ref.\
\cite{Bostwick0,Bostwick}. Also, the cML and the nML$_{\text{Au}}$
have a similar linewidth. The main difference between the two is
the position of the plasmon peak which is determined by the
position of the Dirac point and hence the doping level. The
pML$_{\text{Au}}$, however, has a much lower linewidth over the
whole range of energies indicating a reduced electron-electron scattering.
As the Fermi surface for the pML$_{\text{Au}}$ is rather small (see Fig.\ \ref{figure1} b) the electron-phonon contribution to the linewidth is negligible.
The local maximum in linewidth around $-1$\,eV initial state energy for the pML$_{\text{Au}}$ is not
located at the Dirac point. Therefore, we do not interpret this as
originating from plasmons within the graphene layer according to
\cite{Bostwick0,Bostwick}. We tentatively attribute this peak
to plasmons localized mainly in the Au clusters on top of the
pML$_{\text{Au}}$ that interact with the electrons in the graphene
layer. Varykhalov et al. \cite{Varykhalov1} found a similar feature for graphene/Au/Ni(111) which they attributed to an interaction between Au and graphene. The overall much smaller linewidth for the pML$_{\text{Au}}$
corroborates the conclusion from LEED that the pML$_{\text{Au}}$
is decoupled from the substrate. As mentioned before, the measured linewidth for the pML$_{\text{Au}}$ near the Fermi level is mainly determined by the experimental momentum resolution of $\Delta k=0.023$\,\AA$^{-1}$. This allows us to estimate a lower limit for the carrier lifetime using $\tau=\hbar/(\hbar v_F \Delta k)$. With $\hbar v_F=7.06$\,eV\AA, we find that $\tau > 4$\,fs which is the same order of magnitude as the value reported for multilayer graphene on the C-face of SiC \cite{Sprinkle}.

To gain a deeper insight into the structure of the pML$_{\text{Au}}$ and nML$_{\text{Au}}$, we
measured the Au $4f$ core level spectra using a photon energy of
150\,eV. The spectra in Fig.\ \ref{figure2} b) for the pML$_{\text{Au}}$
(blue) and the nML$_{\text{Au}}$ (green) show two different contributions
to the Au $4f$ core level. The doublet at higher binding energy
was attributed to Au-Si bonds before \cite{Virojanadara,Stoltz}.
The doublet at lower binding energy belongs to Au-Au bonds
\cite{book}. The peak positions are summarized in Table
\ref{table}. 

Combining these observations with the band structures
in Fig.\ \ref{figure1}, we can deduce a schematic as depicted
in Fig.\ \ref{figure2} c). The appearance of a linear dispersion
typical for graphene implies that the C-Si bonds between ZL and
substrate break and a completely sp$^2$-hybridized carbon
monolayer is created. The core level spectra show the existence of
Au-Si bonds for both the nML$_{\text{Au}}$ and the
pML$_{\text{Au}}$. We conclude that the Au atoms intercalate
between the ZL and the substrate replacing the C-Si bonds by Au-Si
bonds. From the core level peak intensity for the
nML$_{\text{Au}}$, we find about one third monolayer of Au
intercalated (one monolayer corresponds to two Au atoms per
graphene unit cell). This is consistent with the observation that
every third carbon atom in the ZL forms a C-Si bond \cite{Emtsev}. For the
pML$_{\text{Au}}$, about one monolayer of gold is intercalated.
From atomic force microscopy (AFM - not shown here) and STM
measurements, we find that additional Au atoms are not
intercalated, but form Au clusters on top of the graphene layer.
Despite the fact that a complete monolayer of gold is intercalated for the pML$_{\text{Au}}$ the substrate does not become metallic. Apart from the graphene bands, there are no other states visible at the Fermi energy. 

The doping behavior for different Au coverages has been addressed
by the theoretical work of Giovannetti et al.\ \cite{Giovannetti}
who predicted p-type doping for graphene on a Au substrate.
Reducing the Au-graphene distance to $d_{\text{AuG}}<3.2$\,{\AA},
however, will lead to n-type doping. The larger amount of
intercalated Au for the pML$_{\text{Au}}$ should increase the
distance between graphene and substrate. This is consistent with
the observed doping behavior as well as the reduced influence of
the ($6\sqrt{3}\times6\sqrt{3}$)R30$^{\circ}$ interface
reconstruction on the Fermi surface and the LEED images of the
pML$_{\text{Au}}$.

\begin{figure}
  \includegraphics[width = 1\columnwidth]{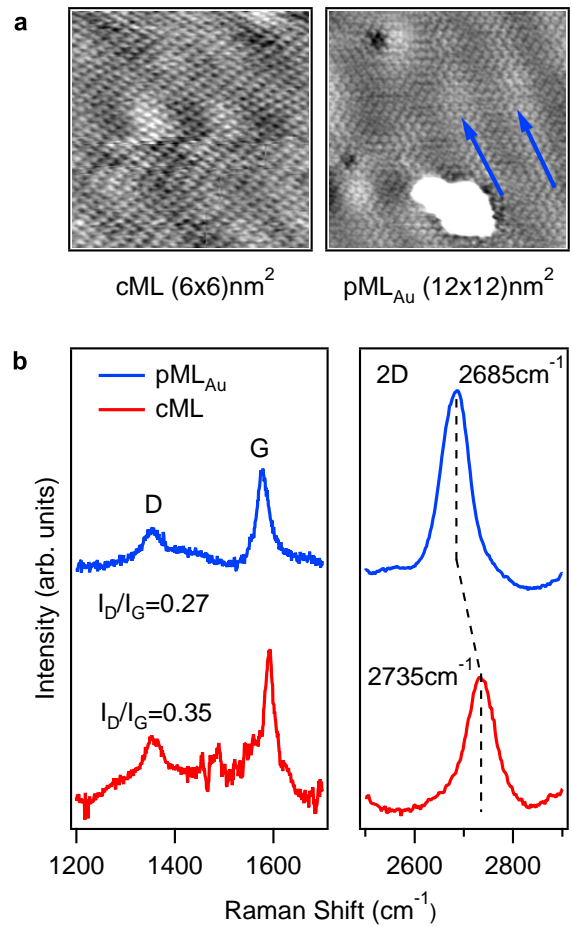}
  \caption{(color online) STM images and Raman spectra reveal improved
  crystalline quality of the p-doped Au-intercalated graphene: panel a) shows topographic
  STM images for the conventional graphene monolayer (left) and the p-doped graphene monolayer intercalated with Au (right). The images
  for the conventional monolayer and p-doped monolayer were recorded at a tunneling current of
  0.2\,nA and a bias of $-0.5$\,V and $-0.4$\,V, respectively. Panel b)
  compares the Raman scattering results for the conventional (red) and the p-doped (blue) graphene monolayer.}
  \label{figure3}
\end{figure}

The peak position for the Au $4f$ doublet
associated with Au-Si bonds shifts by about 860\,meV from
nML$_{\text{Au}}$ to pML$_{\text{Au}}$. This can be related to the observed
difference in the doping and a small change of the work function.
The Au-Au component, on the other hand, shifts only by about
520\,meV. We attribute the Au-Au bonds to Au clusters on top of
the graphene layer. These clusters have an average size of a few
nanometers (see Fig.\ \ref{figure3} a). For such nanoparticles the position of the core levels
depends rather sensitively on the size of the particle
\cite{Boyen,Torelli}. Thus, the shift of the Au-Au component of
the Au $4f$ core level is most likely related to the size of the
particular Au clusters.

As both LEED and ARPES average over a rather large area on the
sample surface, we used STM to gain access to the structure of the
surface on an atomic scale. Fig.\ \ref{figure3} a) shows
topographic images of the cML and the pML$_{\text{Au}}$. The
cML shows a honeycomb lattice with a
($6\sqrt{3}\times6\sqrt{3}$)R30$^{\circ}$ modulation imposed by
the ZL. The graphene lattice of the pML$_{\text{Au}}$ is well
ordered with single defects (black) and some gold clusters
(white). The pML$_{\text{Au}}$ shows a superstructure of parallel
stripes with a width of about 3\,nm as marked by blue arrows in
the right panel of Fig.\ \ref{figure3} a). This superstructure
could be of similar origin as the one reported in \cite{Premlal}
despite the fact that the samples in \cite{Premlal} were prepared
by depositing Au on a cML. The change of the lattice constant of the superstructure between cML and pML$_{\text{Au}}$ is also visible in LEED measurements (see EPAPS).

To further investigate the degree of decoupling of the 
pML$_{\text{Au}}$, Fig.\ \ref{figure3} b) shows Raman scattering data measured 
for the cML and the pML$_{\text{Au}}$. The substrate contribution 
to the Raman data was subtracted from the spectra so that the 
graphene peaks are clearly visible \cite{Lee}. The Raman spectrum for the ZL (not shown here) does not show any graphene related features. The 2D peak of the 
pML$_{\text{Au}}$ (blue) appears at 2685 cm$^{-1}$. It is 
red-shifted by 50 cm$^{-1}$ as compared to the 2D peak of the cML. 
As the 2D peak position is only weakly dependent on charge 
doping \cite{Yan}, we attribute the shift of the 2D peak to an 
increase of the lattice constant in agreement with the LEED data (see EPAPS). The compressive strain present 
in the cML is apparently released in the pML$_{\text{Au}}$. This 
confirms the strongly reduced interactions observed in the analysis 
of the ARPES linewidth. The data in Fig.\ \ref{figure3} b) also suggest that the 
D:G peak intensity ratio has decreased for the pML$_{\text{Au}}$ 
(blue). The D peak only exists in the presence of defects in the 
graphene lattice. A reduced D:G peak intensity ratio therefore 
indicates an improved crystalline quality.


We have shown that it is possible to decouple the graphene ZL
formed on the Si-face of SiC from the substrate by Au
intercalation. This new slightly p-doped graphene has an improved
quality and is only weakly influenced by the underlying substrate.
Our ARPES measurements for the pML$_{\text{Au}}$ reveal a considerable
reduction in linewidth. Our estimation for the carrier lifetime is of the same order of magnitude as the value for multilayer graphene on the C-face of SiC. Therefore, we expect
a considerable increase in carrier mobility for the pML$_{\text{Au}}$ and correspondingly the transport properties of our pML$_{\text{Au}}$ to be closer to those for multilayer graphene on the C-face of SiC.

The authors thank C.\ L.\ Frewin, C.\ Locke and S.\ E.\ Saddow of
the University of South Florida for hydrogen etching of the SiC
substrates. C.\ R.\ A.\ acknowledges funding by the
Emmy-Noether-Program of the Deutsche Forschungsgemeinschaft (DFG).
This work is based in part upon research conducted at the
Synchrotron Radiation Center of the University of
Wisconsin-Madison which is funded by the National Science
Foundation under Award No DMR-0537588.

{\bf EPAPS Available:} Details about the sample preparation, the different experimental techniques and the data analysis are available as EPAPS.

\section*{EPAPS}

\subsection*{Sample preparation}

We have grown graphene on the Si-face of SiC. Our 4H SiC wafers
were hydrogen-etched before insertion into ultra high vacuum
(UHV). To remove residual oxygen impurities we deposited Si from a
commercial electron beam evaporator at a substrate temperature of
800$^{\circ}$C until a sharp (3$\times$3) low energy electron diffraction (LEED) pattern was
observed. We graphitized the samples by direct current heating at
elevated temperature. The sample temperature was measured with an
optical pyrometer at an emissivity of 63$\%$. An annealing
temperature of 1100$^{\circ}$C for five minutes is sufficient for the
formation of the zero layer (ZL), a pure carbon layer, where every
third C-atom forms a chemical bond to a Si-atom in the layer
below. This ZL has no graphene properties, in particular, instead
of the linear dispersion at the $\overline{\mbox{K}}$-point of the
surface Brillouin zone there are two non-dispersing bands at $-0.3$\,eV
and $-1.2$\,eV initial state energy. Upon further annealing at
1150$^{\circ}$C for five minutes a purely sp$^2$-hybridized carbon
layer forms on top of the ZL which shows the linear band structure
characteristic of massless charge carriers in graphene. This
graphene layer is referred to as the conventional monolayer (cML)
in the following.

We deposited gold from a commercial Knudsen cell at room
temperature on a graphene ZL and annealed the sample at
800$^{\circ}$C for five minutes. After this annealing step the
linear dispersion characteristic for graphene is clearly visible
around the $\overline{\mbox{K}}$-point of the surface Brillouin zone.

\subsection*{Photoemission experiments}

The angle-resolved photoemission spectroscopy (ARPES) measurements in Fig.\ 1 a) of the manuscript were done with a
SPECS HSA 3500 hemispherical analyzer with an energy resolution of
10\,meV and monochromatized He II radiation at room temperature.
The Fermi surfaces in Fig.\ 1 b) of the manuscript were measured at the
Synchrotron Radiation Center (SRC) in Madison/Wisconsin using a
Scienta analyzer with an energy resolution of better than 10\,meV,
a photon energy of $\hbar\omega=52$\,eV at a sample temperature of
100K. The angular resolution of 0.4$^{\circ}$ offers a momentum resolution of 0.023\,\AA$^{-1}$ at the Fermi level. This is comparable to the offset that was subtracted in Fig.\ 2 a) of the manuscript for the cML and the pML$_{\text{Au}}$. The Fermi surfaces for the cML and nML$_{\text{Au}}$ were measured with a step size of 0.25$^{\circ}$ along the $\overline{\Gamma\text{K}}$ direction. As the linewidth for the pML$_{\text{Au}}$ is narrower than for the cML and the nML$_{\text{Au}}$ we had to reduce the stepsize to 0.1$^{\circ}$ to allow for reasonable data fitting.

The core level spectra in Fig.\ 2 b) were also measured at
the SRC using a photon energy of $\hbar\omega=150$\,eV. They were fitted with Lorentzian peaks including
a Shirley background.

\subsection*{Scanning tunneling experiments}

The images in Fig.\ 3 a) of the manuscript were measured with a room
temperature scanning tunneling microscope (STM). The SiC samples
with a ZL or cML on top were transferred to the STM chamber in air.
Annealing of the samples at 800$^{\circ}$C was sufficient to
remove any adsorbates from the surface. Au was deposited {\it in
situ} from a commercial electron beam evaporator. The images for
the cML and pML$_{\text{Au}}$ were recorded at a tunneling current of
0.2\,nA and a bias voltage of $-0.5$\,V and $-0.4$\,V, respectively.

\subsection*{Low energy electron diffraction measurements}

\begin{figure}
  \includegraphics[width = 1\columnwidth]{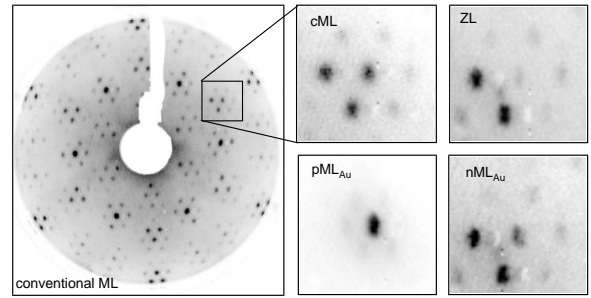}
  \caption{{\bf LEED images} taken at 126\,eV for the conventional graphene monolayer (cML), the zero layer (ZL), the p-doped graphene monolayer (pML$_{\text{Au}}$) and the n-doped graphene monolayer (nML$_{\text{Au}}$). The relative intensity between graphene
  spot and satellite spots is a measure for the strength of the
  substrate influence on the graphene layer. }
  \label{figure_supinf}
\end{figure}

Fig.\ \ref{figure_supinf} shows LEED images recorded at 126\,eV
electron energy. This energy is particularly sensitive to the
graphene coverage [1]. The image for the cML shows the
graphene (10) spot surrounded by satellite peaks from the
($6\sqrt{3}\times6\sqrt{3}$)R30$^{\circ}$ reconstruction. The
graphene (10) spot and the two left lower satellite spots have
roughly the same intensity. For the ZL there is no graphene spot
visible at 126\,eV, only the satellite spots are there. The
pML$_{\text{Au}}$ has a very bright graphene spot, whereas the satellite
peaks are considerably reduced in intensity. Furthermore, the distance between the satellite peaks and the graphene peak is smaller than for the cML indicating a larger lattice constant of the superstructure. This can be related to the strain release in the pML$_{\text{Au}}$ that was revealed in the Raman measurements in Fig.\ 3 b) of the manuscript. The strain release results in a new commensurate periodicity in agreement with the STM measurements in Fig.\ 3 a) of the manuscript that show an increase of the superlattice constant by about a factor of two when comparing cML and pML$_{\text{Au}}$. 
The LEED image for the nML$_{\text{Au}}$ is very
similar to that of the cML indicating a similar influence of the
underlying substrate in both cases.

\subsection*{Raman measurements}

The Raman spectra shown in the manuscript were measured under
ambient conditions using an Argon ion laser with a wavelength of
488\,nm. The laser spot size was 400\,nm in diameter and the laser
power was 4\,mW. The measured graphene signal is rather weak and
superposed by the signal from the SiC substrate. For the Raman
data shown in Fig.\ 3 b) of the manuscript we subtracted the substrate contribution
so that the graphene peaks become clearly visible [2].

The Raman spectra are characterized by three main graphene 
contributions: The G peak corresponds to an in-plane vibration 
of the two sublattices with respect to each other. The D and 
2D peak come from a double resonance scattering process 
[3]. The 2D peak is always visible, whereas the D 
peak only appears in the presence of defects. Both G and 2D
peaks shift as a function of doping [4-6] and
strain [7,8]. Therefore, it is difficult to determine
charge carrier concentration and strain directly from the Raman
data. However, the doping induced shift is strongest for the G
peak [5,6], whereas the effect of strain is more
pronounced for the 2D peak [7]. If the effect of the 
charge carrier concentration can be determined by another procedure 
(in this case ARPES data), the Raman data provide useful information 
about strain.

[1] C. Riedl, A. A. Zakharov and U. Starke, Appl. Phys. Lett. {\bf 93}, 033106 (2008)

[2] D. S. Lee, C. Riedl, B. Krauss, K. von Klitzing, U. Starke and J. H. Smet, Nano Lett. {\bf 8}, 4320 (2008)

[3] S. Reich and C. Thomsen, Phil. Trans. R. Soc. Lond. A {\bf 362}, 2271 (2004)

[4] S. Pisana, M. Lazzeri, C. Casiraghi, K. S. Novoselov, A. K. Geim, A. C. Ferrari and F. Mauri, Nature Mater. {\bf 6}, 198 (2007)

[5] J. Yan, Y. B. Zhang, P. Kim and A. Pinczuk, Phys. Rev. Lett. {\bf 98}, 166802 (2007)

[6] C. Stampfer, F. Molitor, D. Graf, K. Ensslin, A. Jungen, C. Hierold and L. Wirtz, Appl. Phys. Lett. {\bf 91}, 241907 (2007)

[7] M. Huang, H. Yan, C. Chen, D. Song, T. F. Heinz and J. Hone, PNAS {\bf 106}, 7304 (2009)

[8] Z. H. Ni, H. M. Wang, Y. Ma, J. Kasim, Y. H. Wu and Z. X. Shen, ACS Nano {\bf 2}, 1033 (2008)

\end{document}